# Title - Multi-frequency antenna for quasi-isotropic radiator and 6G massive IoT

## Author Information

Bing Xiao[1], Hang Wong[*, 1, 2], Kam Man Shum[1]

1. State Key Laboratory of Terahertz and Millimeter Waves, Department of Electrical Engineering, City University of Hong Kong, Hong Kong, China
2. Shenzhen Research Institute, City University of Hong Kong, Shenzhen, China

## Contributions

B. X. conceived and conducted the research and experiment. H. W. suggested, planned, and supervised the entire study. B. X. and K. M. S carried out the experiment. All authors discussed and reviewed the research and the manuscript.

## Corresponding author

Correspondence to: Hang Wong

## Abstract

An isotropic antenna radiates and receives electromagnetic wave uniformly in magnitude in 3D space. A multi-frequency quasi-isotropic antenna can serve as a practically feasible solution to emulate an ideal multi-frequency isotropic radiator. It is also an essential technology for mobile smart devices for massive IoT in the upcoming 6G. However, ever since the quasi-isotropic antenna was proposed and achieved more than half a century ago, at most two discrete narrow frequency bands can be achieved, because of the significantly increased structural complexity from multi-frequency isotropic radiation. This limitation impedes numerous related electromagnetic experiments and the advances in wireless communication. Here, for the first time, a design method for multi-band (>2) quasi-isotropic antennas is proposed. An exemplified quasi-isotropic antenna with the desired four frequency bands is also presented for demonstration. The measured results validate excellent performance on both electromagnetics and wireless communications for this antenna.

# Introduction

## Introduction

Electromagnetic isotropic radiator is a decades-old problem since it has significance in both theoretical study and practical applications. Early in 1951, Mathis et al. proved that an antenna that radiates equally with a single linear polarization is impossible[1]. However, as the electromagnetic wave has two independent polarization states, their sum can still be uniform in all directions[2]. Since then, numerous quasi-isotropic antennas have been proposed[3]. On the one hand, it serves as a practically feasible solution to emulate an ideal isotropic radiator, which can be used in numerous physical experiments to prove theories and investigate materials' or devices' electromagnetic response. On the other hand, for the upcoming 6th generation mobile network (6G), quasi-isotropic antenna is important for mobile smart devices to be able to connect massive Internet of Things (IoT) terminals in multiple unexpected directions.

IoT will be even more massive in its number of terminals, together with the massive data and computation it brings. Worldwide IoT terminals are expected to increase steadily, reaching 125 billion in 2030[4,5]. Further, the IoT terminal density may even exceed the criteria defined in the 5th generation mobile network (5G) of 1 million per km$^2$ (1 per m$^2$) [6]. Because of this, even though 6G is not finally defined yet, it is expected to support the ubiquitous massive IoT further[7].

Massive IoT is especially embodied in its diverse application scenarios, such as smart home, smart industry, digital agriculture, digital healthcare/medication, and smart traffic[8–13]. Smart mobile devices, such as smart glasses, smartwatches, and unmanned aerial vehicles (UAVs), are supposed to communicate with massive IoT terminals. For example, augmented reality (AR) smart glasses are widely considered the next "big thing" in the evolution of media[14]. Thus, they need an even stronger ability of massive IoT connections than present smartphones, such as communication with smart home infrastructures, low-power wide-area network (LPWAN) terminals in smart cities, the global navigation satellite system (GNSS), and indoor/outdoor positioning beacons.

For mobile devices, their orientations and locations keep changing. The electromagnetic environment is also time-varying. Both factors result in ever-changing wireless channels. However, an overwhelming majority of mobile device antennas are with directional radiation patterns[15] (see Supplementary Note 1). Due to this, the antenna gain in radiation nulls in certain directions approaches zero, in practice, more than 20-30 dB lower than the peak gain of the main beam[16–19]. Thus, if the signal is only received by the antenna's radiation nulls, it is probably lost because the received signal strength (RSS) is lower than the threshold.

Further, even though the wireless signals are well received by the antenna, there is still a difference between signals by line-of-sight (LOS) paths or non-line-of-sight (NLOS) paths. For stable and reliable communication and positioning, we should try our best to guarantee LOS channels while absolutely avoid losing both LOS and NLOS channels (see Supplementary Note 2).

In order to address this problem, researchers developed beam-steering antennas, which direct the main beam to guarantee the wireless channels from un-predefined directions. A beam-steering antenna can be either a single reconfigurable antenna or a phased antenna array[20,21]. Especially for the latter one, it has a higher gain and is widely applied in the enhanced mobile broadband (eMBB) scenarios of 5G. However, unlike eMBB, most IoT scenarios, and thus their protocols, are based on a low data rate, such as tens of kbps[6]. Thus, high-gain beam-steering is not necessary but substantially increases power consumption and the size of the antenna system. Moreover, in the future, the large number of IoT terminals will finally exceed the capability of beam steering of small antennas. For such reasons, beam-steering antenna is not a good choice[22].

In this condition, quasi-isotropic radiated antenna is promising. It has a radiation pattern of an approximate sphere, connecting massive IoT devices in different directions stably and simultaneously. It has a near 0 dBi peak realized gain but can satisfy low-data-rate IoT applications.

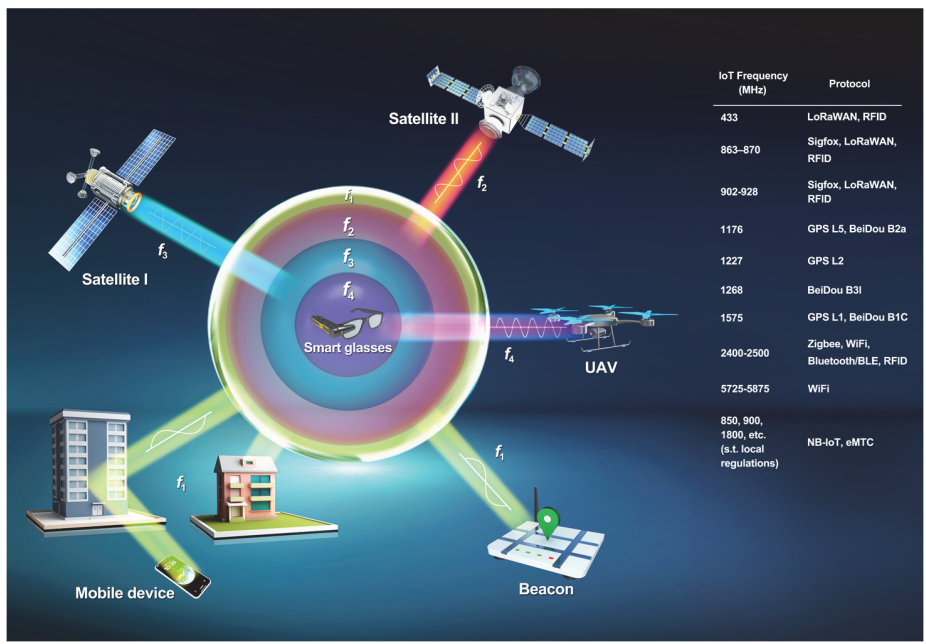

**Fig. 1** Multi-band isotropic antenna applications demonstrated by smart glasses. The smart glasses are equipped with the proposed multi-band quasi-isotropic antenna, being able to communicate simultaneously with UAVs, satellites, mobile devices, and positioning beacons by LOS or NLOS channels. The communications operate on different frequencies.

On the other hand, massive IoT must mean multiple operating frequency bands. The frequency selection is based on communication range, power consumption, network congestion, cost, data rate, etc. IoT usually utilizes narrow-band transmission to balance communication capacity, battery life, and coverage[23], such as 433, 868, 915, 2450, 5800 MHz industrial, scientific, and medical (ISM) bands, and 2/3/4/5G cellular bands (different subject to local regulations)[22]. Moreover, mobile smart devices require the function of outdoor positioning, so the antenna needs frequency bands of GNSS or even multi-GNSS. Multi-GNSS refers to using multiple global navigation satellite systems for precise positioning, such as GPS, BeiDou, and Galileo. Adopting more satellites or signal channels increases the positioning accuracy and improves the reliability and convergence time of the GNSS[24]. Moreover, for indoor positioning, smart mobile devices need to communicate with positioning beacons based on techniques such as Wi-Fi and Bluetooth Low Energy (BLE)[25]. Fig. 1 exhibits part of the multi-band isotropic antenna applications in massive IoT, demonstrated by smart glasses as the subject of communication.

Nevertheless, ever since the quasi-isotropic antenna was proposed and achieved more than half a century ago, at most only two discrete narrow frequency bands can be achieved[3]. It is much far from satisfying future massive IoT in 6G. Multi-frequency simultaneously isotropic radiation/absorption is also in need in numerous physical experiments. Thus, it is listed as one of the most challenging and crucial problems of isotropic antennas[3].

The development history of quasi-isotropic antennas started by combining multiple radiators[26–33] or multiple modes of one radiator[34–43] to supplement the radiation nulls of the single mode. For this mechanism, the isotropic antennas' frequency bandwidth, covering both impedance and gain-deviation bandwidth (see Methods: Gain/directivity deviation), is usually less than 10%. Its robustness is also weak since one isotropic-radiated resonant frequency is produced by at least two resonant modes, whose respective resonant frequencies should be consistent. Slight inconsistencies caused by design or fabrication will significantly degrade the isotropy[3]. When it refers to multi-band isotropic radiation, it also has a limitation when more than two discrete frequencies are considered, because this kind of supplementary structures starts to be quite complex and difficult to achieve.

Another research route is the U-shaped quarter-wavelength ($\lambda$/4) radiator. Since 1994, H. Matzner et al. conducted a series of research on this structure and proved that it can achieve perfect isotropic radiation when the radiated power infinitely approaches zero[44,2,45]. It is a single mode produced by a single radiator that can achieve quasi-isotropic radiation. However, this research only focused on the resonant mode of this structure without considering the feeding arrangement for a practical antenna. Subsequently, researchers found that when this U-radiator is fed in the middle, it suffers from an intrinsically low input resistance and inductive reactance, resulting in difficulty in impedance matching to a standard 50-Ω port[46]. Fortunately, it is easy to match tag chips because the chips' input impedances are small and capacitive[47]. Chihyun et al. applied this U-radiator to a 914-MHz Radio Frequency Identification (RFID) antenna[48]. This antenna matches a commercial tag chip whose input impedance is about 13-j135 Ω. Its feeding port is shorted by a parallel line to tune the impedance matching. Also, for 914-MHz chips, similar U-radiators with or without a small slit on the short-circuit line are proposed[49]. Qing et al. also adopted a short-circuited port to excite this U-radiator. Here, the line-shaped radiator is replaced by a circular-sector-shaped radiator[50]. Its closed-form formulas can be derived accordingly. Recently, Ren et al. adopted a stacked dual-layer instead of a single-layer U-radiator for the typical 50 Ω impedance matching. They also propose a small-loop-fed single-layer U-shaped antenna for tag chips[46]. However, the above-mentioned U-radiator quasi-isotropic antennas can cover only one single narrow frequency band with less than 5% fractional bandwidth.

Here, we demonstrate a method to design quasi-isotropic antennas with > 2 discrete frequency bands for isotropic radiator emulation and for massive IoT. This method adopts $n$ combined U-radiators for $2n$ isotropy frequencies. We will first analyze and design the radiator without considering the feeding port, then introduce the feeding method. Both steps are analyzed from the view of characteristic mode[51]. The exemplified quasi-isotropic antenna covers four frequency bands: 868 MHz (ISM/IoT), 1176 MHz (GPS L5), 1575 MHz (GPS L1), and 2450 MHz (ISM/Bluetooth/BLE/Wi-Fi). They are all currently widely used IoT and GNSS frequency bands. The measured gain deviations of the fabricated antenna in the desired frequency bands are below 5.1 dB.

# Results

**Principle of U-radiator and its transformation.** In the quasi-isotropic radiated U-shaped radiator shown in Fig. 2a, the lengths of the two parallel arms are $\lambda/4$, while $h$ defines the distance between the two paralleled arms. It is also the length of the bottom part of the U-radiator. The geometric center of the U-radiator is at the coordinate origin. If the peak current on the radiator is $I$, the magnitude of the current density $J$ can be written as:

$$J(r) = I[-\delta(x-h/2)\delta(y)\cos(kz+\pi/4)\hat{z} + \delta(y)\delta(kz+\pi/4)\hat{x} + \delta(x+h/2)\delta(y)\cos(kz+\pi/4)\hat{z}]$$
(1)
$$(-\lambda/8 \leq z \leq \lambda/8, \ -h/2 \leq x \leq h/2)$$

For an observer at an angle $(\theta, \varphi)$, the time-averaged, far-zone radiation pattern with the time-harmonic current density can be calculated in Gaussian units as[2,52]:

$$\frac{dP}{d\Omega} = \frac{\omega^2}{8\pi c^3}\left|\hat{\mathbf{k}}\times\left[\hat{\mathbf{k}}\times\int\mathbf{J}(\mathbf{r})e^{i\hat{\mathbf{k}}\cdot\mathbf{r}}dV\right]\right|^2 = \frac{\omega^2}{8\pi c^3}\cdot I^2 \cdot \frac{\sin^2[(k/2)h\sin\theta\cos\phi]}{[(k/2)\sin\theta\cos\phi]^2} \approx \frac{\omega^2}{8\pi c^3}\cdot I^2 h^2$$
(2)

$\hat{k}$ is the unit wave vector. The radiation vanishes in the limit of $h \to 0$ and increases with $h/\lambda$ for a finite and small $h$[2,44,52]. On the other hand, the gain deviation is roughly $1-\sin^2(kh/2)/(kh/2)^2$ from isotropy[52]. The radiation pattern is perfectly isotropic in the limit of $h \to 0$. The gain deviation increases with $h/\lambda$. Based on the above reason, the selection of $h$ should be balanced for the best overall performance of the antenna.

The radiation mechanism of the U-radiator can also be explained qualitatively. In the condition of $h$ approaching 0, as shown in Fig. 2b, the bottom of the U-radiator is the main radiating part. It is an equivalent *ideal dipole* with a nearly evenly distributed current $I$, radiating a donut-shaped pattern with nearly 90° half-power beamwidth (HPBW)[53]. However, for the two parallel arms of the U-radiator, their currents are the same with each other in amplitude along the z-axis, but opposite in direction. Thus, most of the two arms' radiating power is canceled out. The uncanceled power takes only a small portion, resulting from the phase shift and radiation intensity attenuation caused by *h*. Its radiation pattern complements the two radiation nulls of the ideal dipole produced by the bottom of the U-radiator, achieving a combined isotropic radiation.

U-shaped isotropic antennas suffer from a narrow frequency band. The primary reason is the high quality factor (Q) of its fundamental mode due to the above-mentioned small effective radiating area. $Q = \omega_0/BW$, proportional to the angular frequency $\omega_0$ and inversely proportional to the bandwidth $BW$[54]. Antenna's quality factor is lower bounded (LB) by its physical size[55], and $Q_{LB} = 1/ka + 1/(ka)^3$, where $k$ is the wavenumber, and *a* is the radius of the smallest sphere that encloses the antenna. Generally, the lowest quality factor is achieved when the permittivity is the lowest, the aspect ratio is closest to unity, and the field is evenly distributed throughout its volume or surface[56].

Given this, as the bottom of the U-radiator is the main radiator, we expand it to a rhombus (here a rotated square) for a 1:1 aspect ratio, as shown in Fig. 2c. Thus, the eigencurrent distributes more broadly on the main radiator. A low-permittivity substrate Rogers RT5880 will be adopted with relative permittivity $\varepsilon_r$ = 2.2, loss tangent $\tan\delta$ = 0.0009, and thickness 0.787 mm. Moreover, another critical reason for this transformation is the impedance matching when feeding this radiator. This point will be analyzed in detail in the following section, *Capacitive Coupling Feeding Structure (CCFS)*. By the abovementioned methods, the U-radiator can have a lower quality factor, corresponding to a broader frequency bandwidth.

**U-radiator with large inductors embedded.** The present literature found that loading inductors/capacitors or their equivalent structures can pick out and tune specific modes of the radiator[57,58]. For example, if an inductor is embedded in series with the radiator and at the eigencurrent maximum of a specific mode, it will decrease this mode's resonant frequency. On the contrary, if the inductor is at an eigencurrent null, the resonant frequency will remain unchanged. Here, we will adopt this method to produce dual isotropic radiated modes from a single U-radiator.

The results in the following two sections are analyzed and simulated on perfect electric conductor (PEC) models by the CMA (characteristic mode analysis) solver of 3D electromagnetic simulation software CST[59] without any feeding sources (see Methods: The theory of characteristic mode (TCM)). The aim is to investigate the radiation principle.

We embed two $L = 40$ nH inductors near each end of the U-radiator. The locations are the eigencurrent maximum of the 3$^{rd}$ order mode, while the eigencurrent of the 1$^{st}$ order mode is quite small here. Fig. 2d compares the characteristic angles (CAs) of the U-radiator's characteristic modes before and after embedding inductors. The resonant frequency of a given mode corresponds to its curve's intersection with the $CA = 180°$ line. The first four modes (modes with the lowest four resonant frequencies) are exhibited. We can see that the resonant frequencies of the 1$^{st}$ and 3$^{rd}$ order modes, $f_1$ and $f_3$, all decrease after embedding inductors. However, the decrease of $f_3$ is much greater than $f_1$.

Fig. 2f shows the corresponding eigencurrent distribution and radiation patterns before and after embedding inductors. Before adding the inductors, only the 1$^{st}$ order mode (fundamental mode) radiates isotropically, resonating at $f_1$, as we have already discussed. The 2$^{nd}$ and 4$^{th}$ order modes exhibit co-directional current on their arm pairs. For both conditions, the U-radiator retreats to a common dipole with its 1$^{st}$ or 2$^{nd}$ order mode in nature[57]. They are all not isotropically radiated.

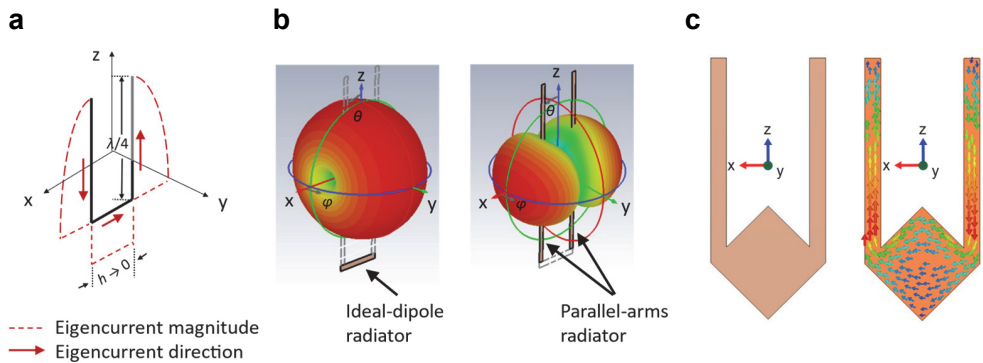

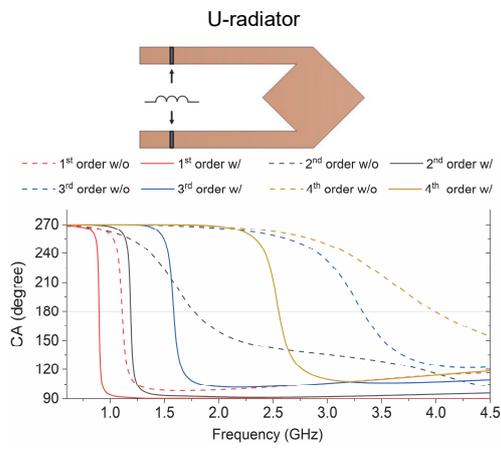
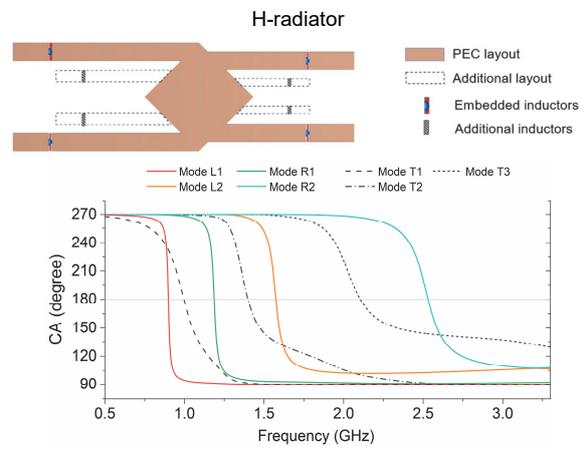

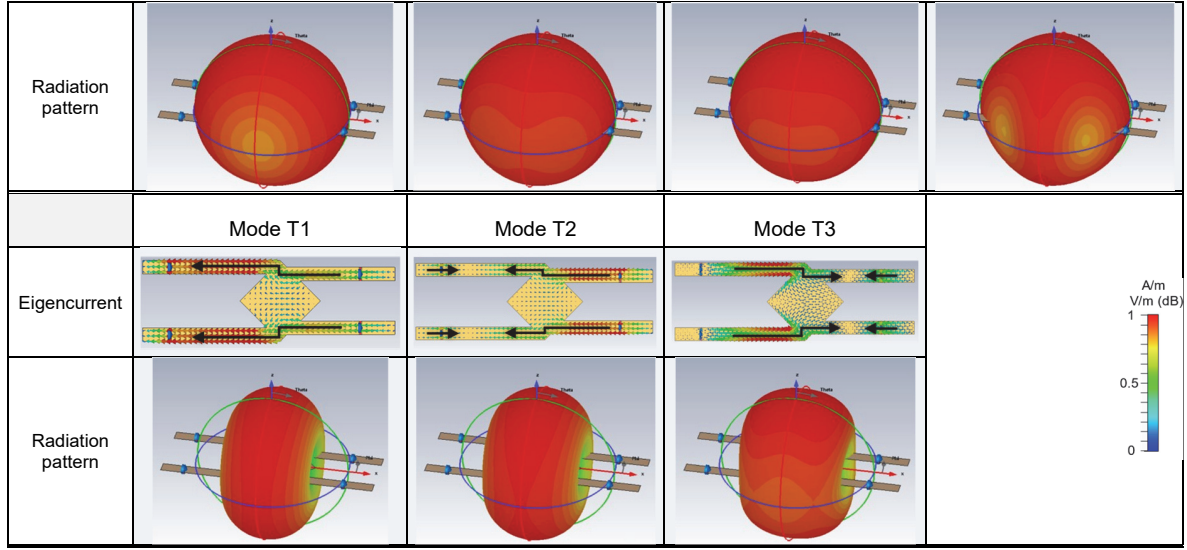

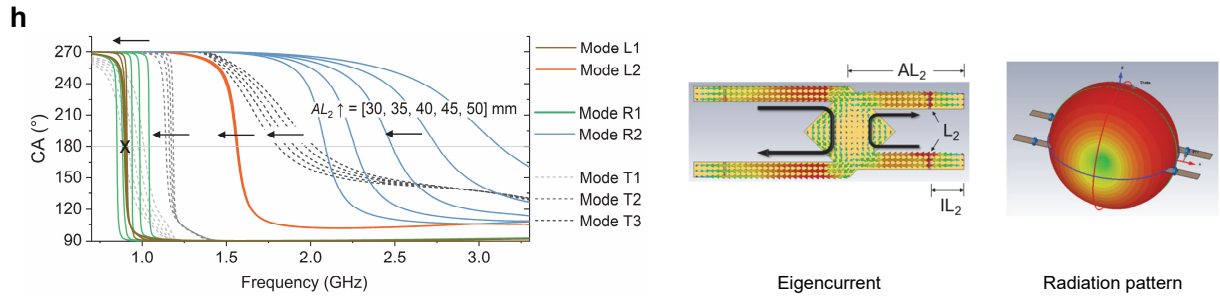

Fig. 2. Evolution from a simple U-radiator to an inductor-embedded quad-isotropy-mode H-radiator. **a** Eigencurrent distribution of the fundamental mode of U-radiator in the y = 0 plane. **b** Conceptual schematic of the radiation patterns produced by the bottom part and the parallel arms, respectively. **c** Transformed U-radiator and its eigencurrent distribution. **d** U-radiator and its characteristic angles (CAs) without and with inductors. **e** H-radiator by combining two U-radiators and its resulting CAs, together with possible additions to the existing layout for more isotropy frequencies. **f** Normalized eigencurrent distribution and radiation patterns of the U-radiator without and with embedded inductors. **g** Normalized eigencurrent distributions and radiation patterns of the H-radiator, including four isotropy modes and three non-isotropy modes. **h** A counter-example demonstrates isotropy deterioration when $AL_2$ = 45 mm, presented by this mode's CA variation, normalized eigencurrent distribution, and radiation pattern.

For the 3$^{rd}$ order mode, it resonates at $f_3 \approx 3f_1$, correspondingly $\lambda_3 \approx \lambda_1/3$. The currents on the arm pair are in the opposite direction, thus mostly canceling out, similar to the 1$^{st}$ order mode. However, as we have set $h/\lambda_1 \approx 0.1$ for the balance of antenna efficiency and isotropy at $f_1$, $h/\lambda_3$ will be 0.3. This large ratio significantly increases the directivity deviation (see Methods: Gain/directivity deviation), deteriorating the isotropy of the U-radiator. That is the reason for the 3$^{rd}$ order mode inherently not radiating isotropically.

However, the 3$^{rd}$ order mode changes after we embed the inductors. Inductors at those specific locations extend the equivalent electrical length of the U-radiator, slightly for the 1$^{st}$ order mode, but significantly for the 3$^{rd}$ order mode. Thus, they decrease $f_1$ slightly, but $f_3$ significantly. It also means $h/\lambda_3$ is greatly decreased, while $h/\lambda_1$ is quite less influenced. This measure keeps both $h/\lambda_1$ and $h/\lambda_3$ in an acceptable range simultaneously. As a result, the directivity deviation of the 3$^{rd}$ order mode decreases from 8.7 dB to 3.6 dB. The resonant frequency decreases from 3.32 GHz to 1.59 GHz. By this method, the 3$^{rd}$ order mode also turns to an isotropy mode. Thus, dual isotropic modes are achieved in one single U-radiator, as shown in Fig. 2f.

**Combine multiple U-radiators.** Further, multiple U-radiators can be combined for more isotropy frequencies. Here, we combine two dual-mode dual-frequency U-radiators to form an H-shaped radiator for the desired four isotropic frequencies, as shown in Fig. 2e (see Supplementary Figure 1 for detailed dimensions), together with the CAs of the first seven modes of this new H-radiator. Among them, there are four isotropy modes. Mode L1 and its 3$^{rd}$ order harmonic, Mode L2, both resonate mainly on the left arm pair. Mode R1 and its 3$^{rd}$ order harmonic, Mode R2, both resonate mainly on the right arm pair, as the simulated eigencurrent distribution and radiation patterns shown in Fig. 2g.

Notably, the combination of U-radiators also introduces three new modes, Modes T1, T2, and T3. Their eigencurrents go horizontally through the top and bottom corners of the central rhombus. Eigencurrents on both left and right arm pairs are enhanced instead of being canceled out. Thus, they are equivalent 0.5$\lambda$ dipole mode, and two 1$\lambda$ dipole modes. However, these modes will not affect the desired isotropy modes, because they will not be excited finally, which will be explained further in the following section, *Capacitive coupling feeding structure (CCFS)*.

Because higher-frequency U-radiators will have shorter arm length *AL* and shorter arm spacing *h*, they can be buried inside lower-frequency U-radiators in the same layer. The possible added arm pairs for more isotropy frequencies are indicated by dashed lines in Fig. 2e. Finally, this schematic can easily achieve 2*n* isotropy frequencies by *n* combined U-radiators with a simple structure.

**Mode purification on isotropy frequencies.** However, the combination of U-radiators mentioned above has one limitation: On each isotropy frequency, only one isotropy or its harmonic modes can be allowed to resonate. When the resonant frequencies of any two of them coincide, the expected isotropy property will deteriorate. Here, we present a counterexample to demonstrate this situation (see Supplementary Figure 2 for detailed dimensions). As shown in Fig. 2h, the length of the right arm pair $AL_2$ varies from 30 mm to 50 mm with five steps. The relative location of the inductors $L_2$ on the right arm pair, $IL_2$, changes its value to keep the inductors at the eigencurrent maximum of Mode R2 during the variation. When $AL_2$ = 45 mm, the resonant frequencies of Modes L1 and R1 coincide. In this condition, the eigencurrents on the left and right arms cancel out each other, leaving only the radiation from the central rhombus. Then, both original isotropy modes retreat to the same short-dipole mode. The axis of the radiation pattern of the dipole mode is parallel to the middle bar of *H*, as presented in Fig. 2h. At the same time, the directivity derivation increases from less than 4 dB to 10.2 dB. This result verifies our proposed qualitative analytical model of the U-radiator, as presented in Fig. 2b.

**Capacitive coupling feeding structure (CCFS).** In the above sections, multiple isotropy eigenmodes can be designed to resonate at desired frequencies. In this section, they will be excited by the proposed CCFS.

We know the U-radiator suffers from an intrinsically low input resistance and inductive reactance, resulting in difficulty in impedance matching to a standard 50-Ω port[46]. In the present literature, U-radiator quasi-isotropic antennas adopt feeding methods by short-circuited port, small loop, etc. Within single <5% fractional bandwidth, their impedance matching capability is enough. However, the desired antenna needs to cover multiple discrete frequency bands, 868 MHz (863–870 MHz), 1176 MHz (1176.45±1.023 MHz), 1575 MHz (1575.42±1.023 MHz), and 2450 MHz (2400–2480 MHz). They are in a frequency span of 863-2480 MHz with 97% fractional bandwidth. Thus, a wideband impedance-matching feeding method by capacitive coupling is proposed for this multi-band matching.

As shown in Fig. 3a, the rhombic main radiator is split along its horizontal diagonal to arrange the feeding port. Then, the input impedance of this transformed center-broadened radiator is observed to be different from a typical thin U-radiator. On the four resonant frequencies, the input impedance's real parts increase to close to 50 Ω, no more as low as those shown in *Introduction*. It is because of the broadened width of the main radiator from a narrow strip to a rhombus. However, the imaginary parts of the input impedance are still positive, indicating they are still inductive, the same with the common thin U-radiator.

Here, we apply a CCFS to balance the imaginary parts of the input impedance at multiple frequencies. Here, CCFS is a parallel plate capacitor containing a rectangular patch with a feedline, the upper half of the H-radiator, and the substrate between them. After the H-radiator's dimensions were determined by characteristic mode analysis, the parameters of CCFS, as shown in Fig. 3a, can be tuned for an optimized equivalent capacitance $C_e$ for multi-frequency impedance matching.

We know the capacitance of a typical parallel plate capacitor $C = \varepsilon S/d$, where $\varepsilon$ is the permittivity of the substrate, $S$ is the size of the plates, and $d$ is the distance between the two plates. Here, the two plates are all irregularly shaped with different areas. The undefined variables are the length *CL* and the width *CW* of the patch, and the length *ML* and the width *MW* of the feedline. After CCFS is inserted in series with the feeding port, we tune the above variables for the input impedances of all four resonant frequencies to move anticlockwise along the constant resistance circle by decreasing $C_e$[60]. Then, they will be closer to the center of the Smith chart, representing the 50 Ω standard port for common smart mobile devices.

For a quantitative study of $C_e$, the CCFS is also simulated by the 3D Electrostatic (Es) Solver of *CST* (see Supplementary Figure 3). Its equivalent capacitance $C_e$ = 2.18 pF. Then, if the CCFS is replaced by a lumped 2.18 pF capacitor, the reflection coefficient result is quite similar, as shown in Fig. 3b.

In order to check the mode purity for good isotropy properties on each resonant frequency, the excited mode proportion is simulated by an indicator of the modal weighting coefficient (MWC) (see Methods: The theory of characteristic mode (TCM)). The simulated results of MWCs in Fig. 3b show that each resonant frequency is produced by a single highly dominant isotropy mode. MWCs of other isotropy and harmonic modes are below 0.2 in the desired frequency bands. This result indicates that the mode purities are good, ensuring good isotropy properties.

Moreover, for eigencurrents of all non-isotropy modes, i.e., Modes T1, T2, and T3 in Fig. 2g, the feeding port is at the minimum of the eigencurrent distributions, causing a near-zero MWC. Thus, these modes are not excited, not affecting the mode purity, thus the isotropy property.

**a**

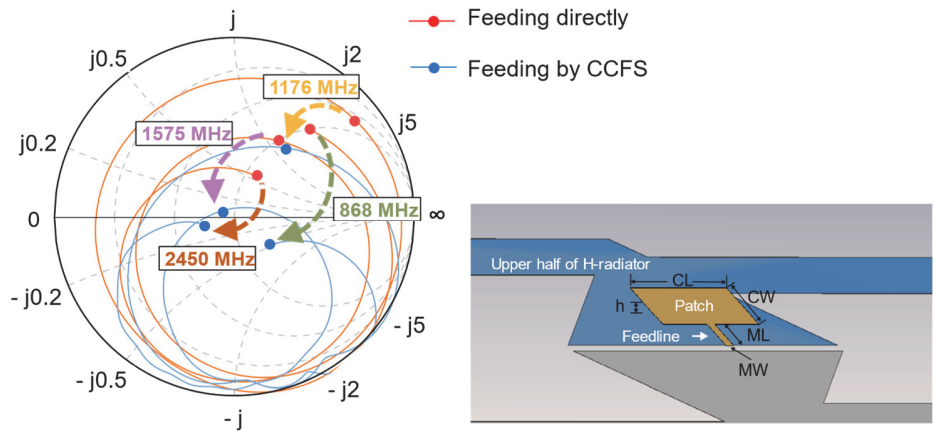

**b**

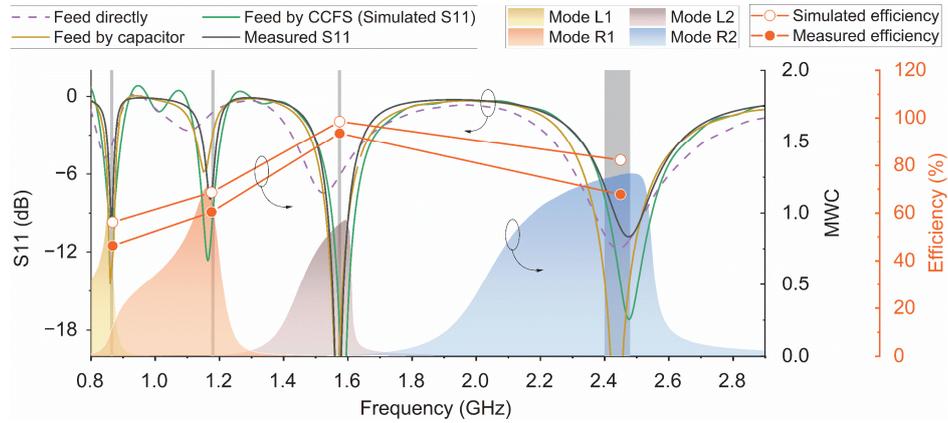

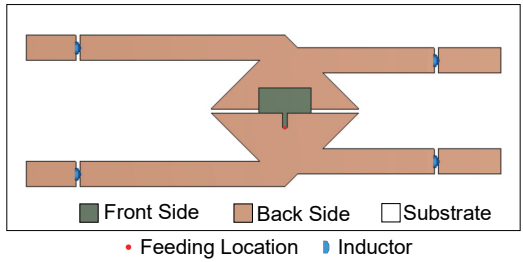
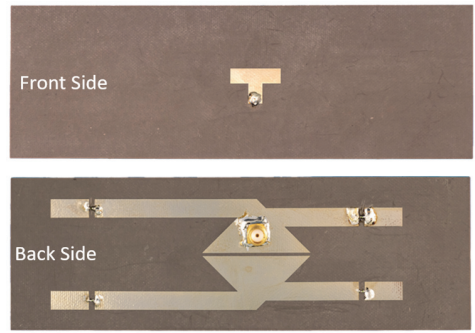

Fig. 3. Excitation and impedance matching by the proposed CCFS. **a** Comparison of CCFS and direct feeding at the four resonant frequencies in Smith Chart. **b** Simulated and measured reflection coefficients ($S_{11}$), MWC, and total efficiency. **c** Schematic of the antenna. **d** Front and back sides of the antenna prototype.

**Measurement of electromagnetic far field and wireless communication.** Finally, the antenna is fabricated on Rogers RT/duroid 5880 laminate. The total size of the antenna is 150 mm × 50 mm × 0.8 mm, corresponding to $0.43\lambda_L \times 0.14\lambda_L \times 0.002\lambda_L$, where $\lambda_L$ is the wavelength of the lowest frequency, 868 MHz. Fig. 3b shows the simulated and measured reflection coefficient ($S_{11}$) and total efficiency of the antenna. The measured reflection coefficient is below -6 dB for 0.857-0.872, 1.160-1.182, 1.523-1.618, and 2.385-2.570 GHz, which can cover the desired IoT and GNSS frequency bands for multi-narrow-band applications. The schematic and the fabricated prototype are shown in Fig. 3c & d. It is measured in a microwave anechoic chamber, *Satimo StarLab* 3D Antenna Measurement System, which is set up as in Fig. 4a. The measured efficiencies are 46.3%, 60.5%, 93.6%, and 67.8%, respectively. Far-field radiation properties are also measured in the chamber. The measured gain deviations at four frequency bands are shown in Fig. 4c. They are 4.2 dB, 4.2 dB, 5.1 dB, and 4.5 dB, respectively. The simulated and measured radiation patterns are in Supplementary Figure 4.

Notably, the 1575 and 2450 MHz frequency bands, achieved by the H-radiator's higher harmonic isotropy modes, have obviously wider bandwidths and higher total efficiencies, compared with 868 and 1176 MHz by the fundamental isotropy modes. At the same time, they keep similar gain deviations. It proves that our proposed higher-order U-radiator isotropy modes can also achieve high-performance quasi-isotropic radiation.

We also built up an indoor wireless communication experiment to further validate the real-world performance of this antenna. The communication link is set up as in Fig. 4b. For the transmitter side on the left, the dual-polarized horn antenna is connected to an arbitrary waveform generator (Agilent M8190A) to transmit the modulated waveforms. The final generated waveforms are continuous waves (CWs) with an input power of 30 dBm and a symbol rate of 1 MBaud. The modulation schemes are QPSK for the two lower frequency bands, and 16 QAM for the two higher frequency bands. Both schemes are widely applied in present IoT communications, such as NB-IoT, Zigbee, and Wi-Fi.

The receiver side on the right is our quasi-isotropic antenna, at the boresight of the horn antenna ~1.5 m away from it. The quasi-isotropic antenna is connected to a vector signal analyzer (VSA, PathWave 89600) to receive and demodulate the received waveforms. The VSA can provide real-time constellation diagrams, eye diagrams, error vector magnitudes (EVMs), etc. Fig. 4d shows the received signals with constellation diagrams in every frequency and cardinal direction (+X, -X, +Y, -Y, +Z, -Z). In each measurement, the EVM is lower than 8.0%, demonstrating that the wireless communication link maintains a high level of signal fidelity and that the antenna meets the requirement of accurate data transmission for IoT and GNSS.

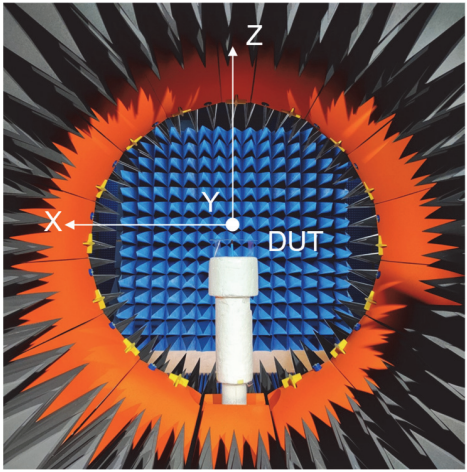
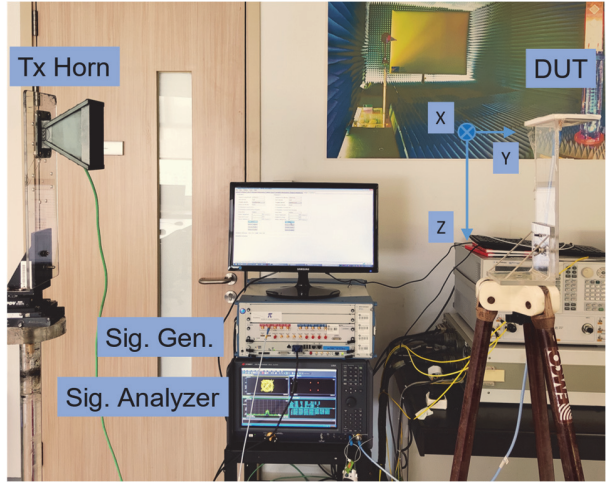
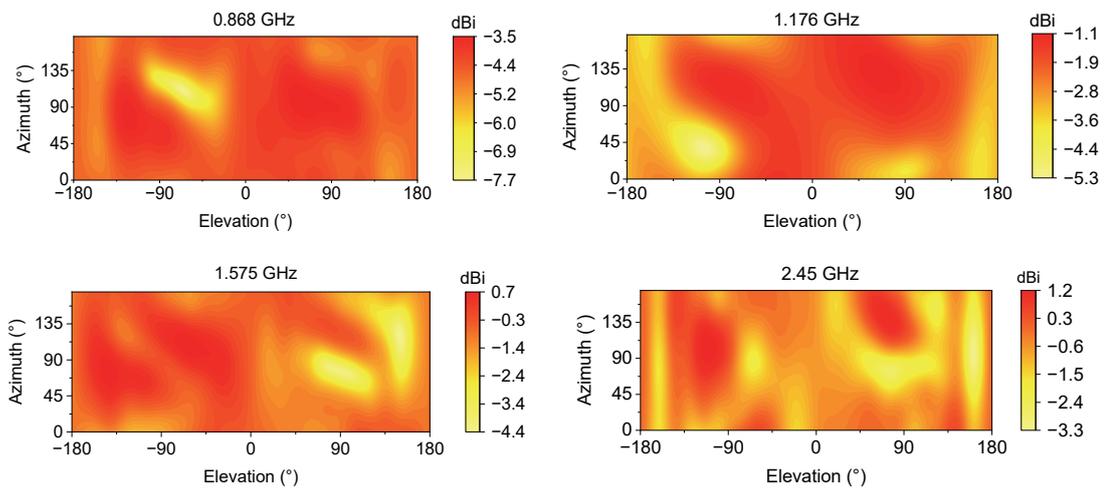
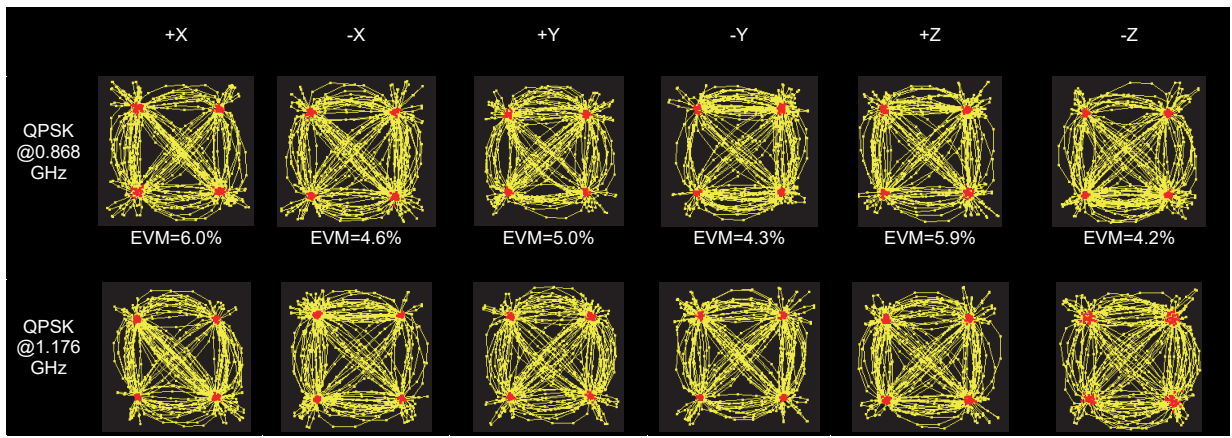

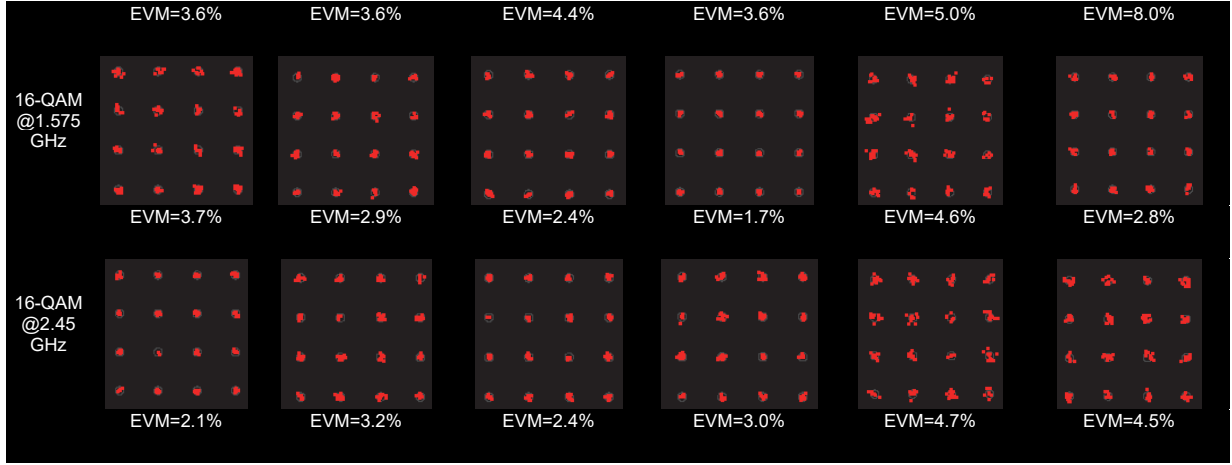

Fig. 4. Measured multi-frequency isotropic radiation characterization of the antenna. **a** Measurement setup in a microwave anechoic chamber for the far-field radiation. **b.** Measurement setup for wireless communication in indoor environment. (The axis represents the antenna's direction. Here, the antenna's -Y direction is pointing to the horn, thus -Y is under testing). **c** Measured gain deviations at the four frequencies. **d** Constellation diagrams and EVMs of the received QPSK or 16QAM modulated signals in every frequency and cardinal directions (+X, -X, +Y, -Y, +Z, -Z)

# Discussion

Multi-frequency quasi-isotropic antenna has significant meaning for electromagnetic isotropic source emulation in physical experiments and future mobile smart devices for massive IoT in 6G. However, it is always difficult to achieve because of the significantly increased geometry complexity after the frequency bands are more than two. Here, we report on an approach to address this problem by novel mode generation and excitation methods, including the rhombic main radiator, embedded large inductors, combined multiple U-radiators, mode purification, and capacitive coupling feeding. For the first time, the quasi-isotropic antenna's frequency bands exceed two and can easily extend to even more. An exemplified quad-band quasi-isotropic antenna is presented with a high-to-low frequency ratio of up to 2.8. The measured gain deviation is below 5.1 dB. The antenna has a quite simple 2D structure in small size with a single substrate layer. These merits promote its applications to mobile smart devices, such as smart glasses, smartwatches, and UAVs, enabling them to connect future high-density massive IoT terminals in 6G.

## Methods

**Gain/directivity deviation.** The antenna gain is defined by $G(\theta, \varphi) = 4\pi \times$ (power radiated per unit solid angle)/(total accepted power). Gain deviation refers to the difference between the maximum and minimum antenna gain in the entire 3D radiation sphere at a certain frequency. An ideal isotropic antenna's gain deviation is zero. Until now, there has not been a uniform standard of the upper limit of gain deviation for a quasi-isotropic antenna. In the present literature, most quasi-isotropic antennas' gain deviations are lower than 10 dB[3]. In this paper's demonstrated quad-band quasi-isotropic antenna, the gain deviations at the four desired frequency bands are below 5.1 dB.

Directivity is defined by $D(\theta, \varphi) = 4\pi \times$ (power radiated per unit solid angle) / (total radiated power). Directivity deviation is similar to gain deviation but describes the difference in directivity, so the *total accepted power* is replaced by the *total radiated power* in the denominator. For characteristic mode analysis without the feeding port, the accepted power of the concerned structure is not available. The directivity deviation of the radiator is calculated instead.

**The theory of characteristic mode (TCM).** TCM can bring physical insight into the radiating phenomena of antennas without or with feeding arrangement[61,62,51,63]. For pure PEC antenna models, characteristic modes only depend on the shape and size of the PEC structure. At a given frequency, it can determine whether or not the *n*-th order characteristic mode is at resonance by its characteristic angle (CA) $\theta_n$, eigenvalue $\lambda_n$ (or $\Lambda_n$), or modal significance $MS_n$. When $\theta_n = 180°$, $\Lambda_n = 0$, or $MS_n = 1$, the mode is at resonance. Eigencurrents are real (all in phase), forming an orthogonal set over the concerned surface. Then, the total current can be expanded into those orthogonal modes. Here, we only adopt the eigencurrent distribution at exactly the resonant frequency of the concerned mode[64].

The modal weighting coefficient (MWC) indicates how strongly a mode is excited. It is expressed as: $MWC = <\boldsymbol{J}_n, \boldsymbol{E}^i>/(1+j\Lambda_n) = \iint_s(\boldsymbol{J}_n \cdot \boldsymbol{E}^i)dS/(1+j\Lambda_n)$. $\boldsymbol{E}^i$ is the impressed electric field on the conductive surface $S$ of the radiator. Thus, $<\boldsymbol{J}_n, \boldsymbol{E}^i>$ indicates the coupling degree between the intrinsic characteristic mode of the radiator and the external incident field from the feeding source. A higher MWC indicates a stronger excitation of a specific mode.

Mainstream 3D electromagnetic simulation software, such as CST[59], provides a CMA (characteristic mode analysis) solver based on TCM. Especially, thanks to the method-of-moment (MoM) algorithm, a much faster result can be obtained for perfect electric conductor (PEC) models, compared with full-wave simulation. This merit benefits the fast analysis of the characteristic mode composition of an arbitrary radiating structure.

## Ethics declarations

### Competing interests

The authors declare no competing interests.

# Supplementary Information

**Supplementary Note 1: Radiation patterns**

Radiation pattern is the radiation properties of an antenna as a function of space coordinates. Due to the principle of reciprocity, reception patterns are the same as radiation patterns for reciprocal antennas. Thus, we will only use the term *radiation patterns* to simplify the expression[1].

**Supplementary Note 2: Line-of-sight (LOS) and non-line-of-sight (NLOS) channels**

For wireless channels, there are both line-of-sight (LOS) channels where the transmitting (Tx) and receiving (Rx) nodes can see each other directly, and non-line-of-sight (NLOS) channels where they cannot[2]. For stable and reliable communication and positioning, we should try our best to guarantee LOS channels, and absolutely avoid losing both LOS and NLOS channels. Based on the experiments of existing research, NLOS is worse than LOS in multiple aspects due to the power loss from scattering and the increased traveling path. First, the data transmission quality of NLOS is worse because of the lowered signal strength[3]. Second, the latency is increased[4]. Third, the positioning based on received signal strength indication (RSSI) or propagation delay is less accurate[4–6]. Below are the different cases in rich and poor scattering environments.

1. In a rich scattering environment, such as indoor scenarios, the LOS signal is often not received[7]. Probably the scattered NLOS signal can still be received. The expenses are the lower Rx power, higher latency, and over-estimated distance for positioning.

2. In a poor scattering environment, mobile devices risk not receiving both LOS and NLOS signals. Its typical scenarios are 1) space near the open ground, 2) intra-communication in a swarm of aircraft in the sky, 3) downlink of aircraft (e.g., from ground stations/beacons to aircraft, or from satellites to aircraft).

Furthermore, from the experiment results, the quasi-isotropic antenna can receive more power than the 2D omni-directional antennas in an NLOS position, such as in the office partition made of metal and plastic[3]. This property is especially preferable for wireless energy harvesting to collect ambient electromagnetic power in the environment, for the power supply of smart mobile devices.

**Supplementary Figure 1: Dimensions of the exemplified H-radiator PEC model**

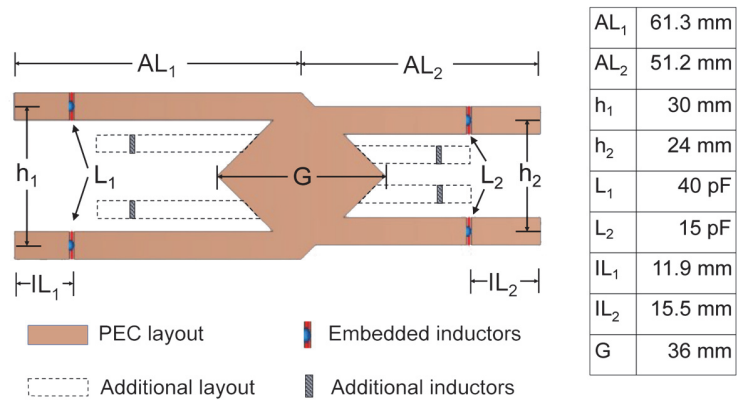

**Supplementary Figure 2: Dimensions of the H-radiator when the resonant frequencies of two isotropy modes (Modes L1 and R1) coincide.**

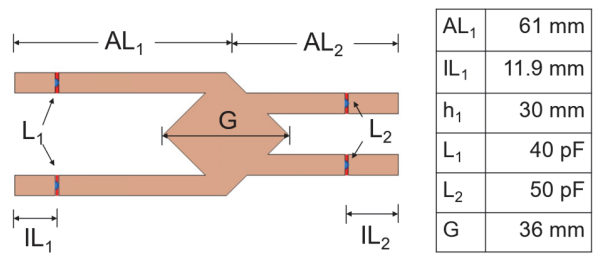

**Supplementary Figure 3: Dimensions of the final fabricated antenna.**

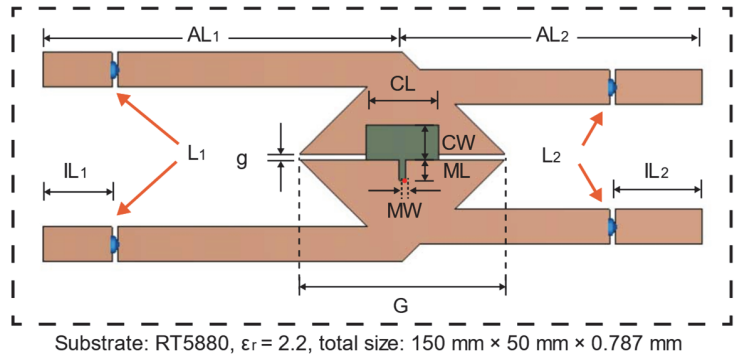

Substrate: RT5880, $\varepsilon_r$ = 2.2, total size: 150 mm × 50 mm × 0.787 mm

Unit: mm

| $AL_1$ | $AL_2$ | $IL_1$ | $IL_2$ | $CL$ | $CW$ | $ML$ | $MW$ | $G$ | $g$ | $L_1$ | $L_2$ |
|---|---|---|---|---|---|---|---|---|---|---|---|
| 66.7 | 47.5 | 13.2 | 13.1 | 16.2 | 4.8 | 6.5 | 4.5 | 36 | 1.0 | 36 (nH) | 20 (nH) |

**Supplementary Figure 4: Simulated and measured radiation patterns of the fabricated antenna at four desired frequency bands.**

a

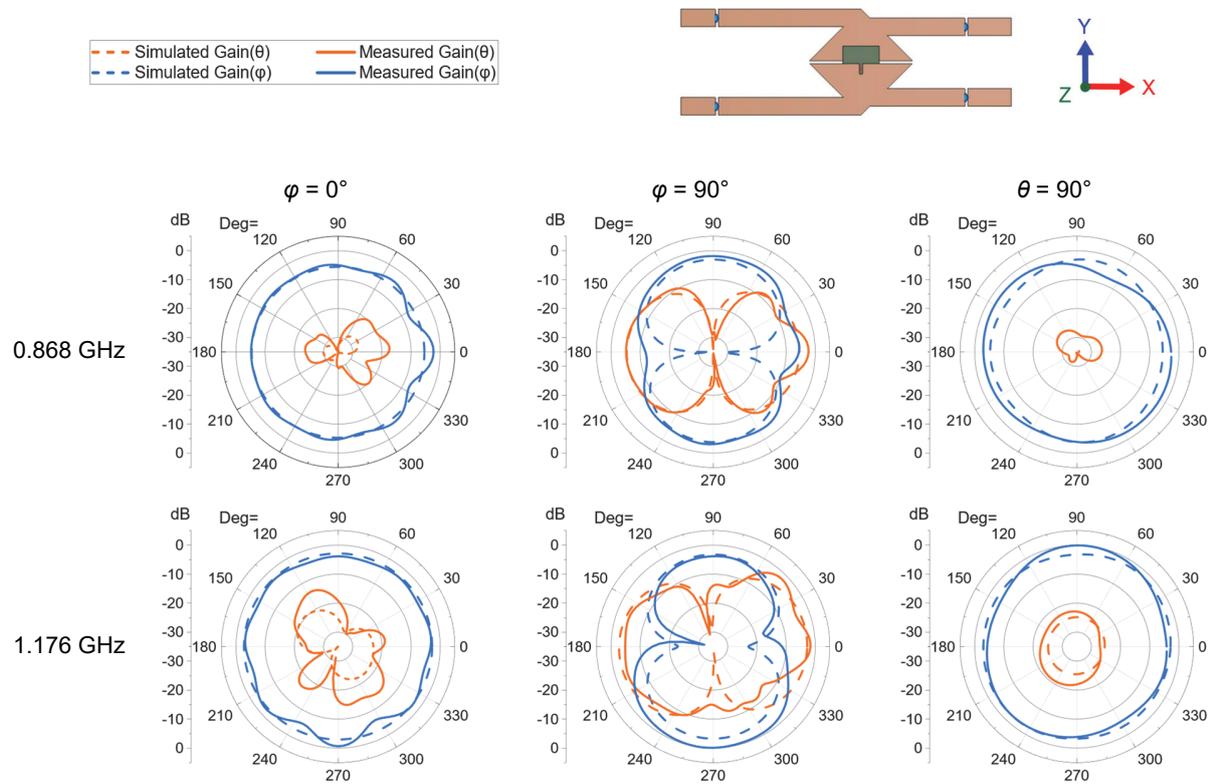

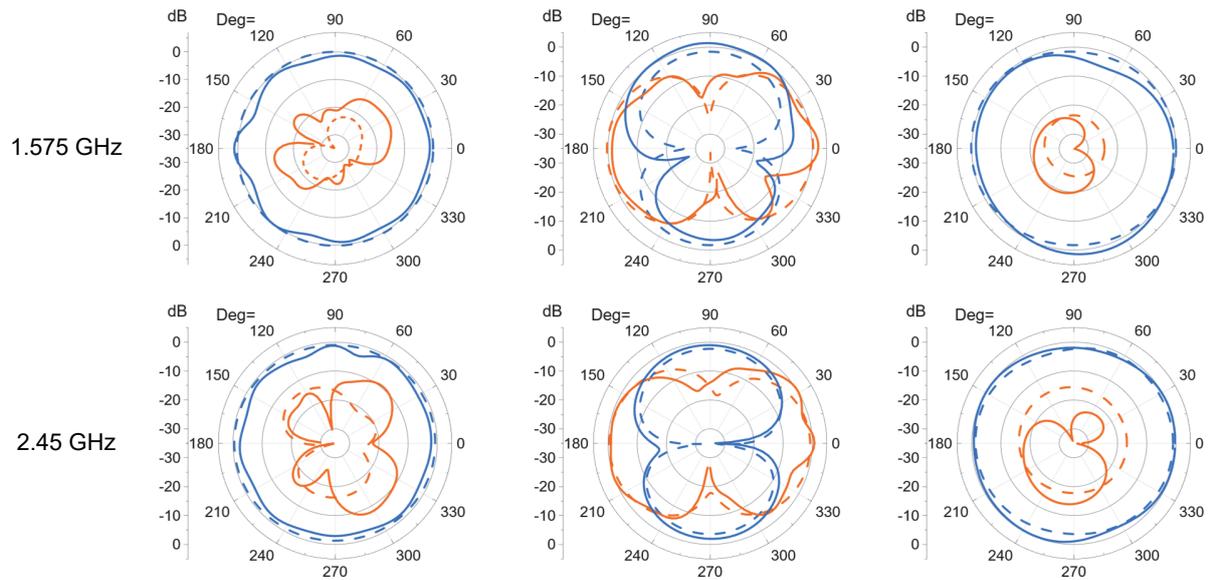